\documentclass[conference]{IEEEtran}
\IEEEoverridecommandlockouts
\usepackage{cite}
\usepackage{amsmath,amssymb,amsfonts}
\usepackage{algorithmic}
\usepackage{graphicx}
\usepackage{textcomp}
\usepackage{xcolor}
\usepackage{hyperref}
\usepackage{graphics}
\usepackage{multirow}
\usepackage{amsthm}
\usepackage[ruled,vlined]{algorithm2e}
\usepackage[compact]{titlesec}
\usepackage{amsmath}
\makeatletter
\newcommand{\removelatexerror}{\let\@latex@error\@gobble}
\makeatother
\def\BibTeX{{\rm B\kern-.05em{\sc i\kern-.025em b}\kern-.08em
    T\kern-.1667em\lower.7ex\hbox{E}\kern-.125emX}}
\newtheorem{definition}{Definition}
\newtheoremstyle{definition} % name
    {\topsep}                    % Space above
    {\topsep}                    % Space below
    {\itshape}                   % Body font
    {}                           % Indent amount
    {\scshape}                   % Theorem head font
    {.}                          % Punctuation after theorem head
    {.5em}                       % Space after theorem head
    {}  % Theorem head spec (can be left empty, meaning ‘normal’)
\makeatletter

\newcommand*\bigcdot{\mathpalette\bigcdot@{.8}}
\newcommand*\bigcdot@[2]{\mathbin{\vcenter{\hbox{\scalebox{#2}{$\m@th#1\bullet$}}}}}
\makeatother

\begin{document}
%\fontsize{9.90}{12}\rm
\setlength{\textfloatsep}{9pt plus 0.0pt minus 5.0pt}
%\titlespacing{\section}{0pt}{2ex}{1ex}
\title{Precision and Fitness in Object-Centric Process Mining\\

\thanks{We thank the Alexander von Humboldt (AvH) Stiftung for supporting our research.}
}

\author{\IEEEauthorblockN{Jan Niklas Adams}
\IEEEauthorblockA{\textit{Chair of Process and Data Science} \\
\textit{RWTH Aachen University}\\
Aachen, Germany \\
niklas.adams@pads.rwth-aachen.de}
\and
\IEEEauthorblockN{Wil M.P. van der Aalst}
\IEEEauthorblockA{\textit{Chair of Process and Data Science} \\
\textit{RWTH Aachen University}\\
Aachen, Germany \\
wvdaalst@pads.rwth-aachen.de}
}

\maketitle
\renewcommand\IEEEkeywordsname{Keywords}
\begin{abstract}
Traditional process mining considers only one single case notion and discovers and analyzes models based on this. 
However, a single case notion is often not a realistic assumption in practice. Multiple case notions might interact and influence each other in a process. Object-centric process mining introduces the techniques and concepts to handle multiple case notions. So far, such event logs have been standardized and novel process model discovery techniques were proposed. However, notions for evaluating the quality of a model are missing. These are necessary to enable future research on improving object-centric discovery and providing an objective evaluation of model quality. In this paper, we introduce a notion for the precision and fitness of an object-centric Petri net with respect to an object-centric event log. We give a formal definition and accompany this with an example. Furthermore, we provide an algorithm to calculate these quality measures. We discuss our precision and fitness notion based on an event log with different models. 
Our precision and fitness notions are an appropriate way to generalize quality measures to the object-centric setting since we are able to consider multiple case notions, their dependencies and their interactions. 
\end{abstract}


\section{Introduction}
In the past years, process mining\cite{ProcessMiningDSIA} has introduced a data-driven way to discover and analyze processes. The introduced techniques and algorithms enable organizations to discover the true underlying process from the data of its execution rather than manually designing what is assumed to be the process.

These techniques work on an event log which records the executed activities for different cases. Each event has an activity and is attributable to exactly one case, e.g., the customer, order, patient, etc.. The attribute used to assign an event to a case is called the case notion. The event log, therefore, describes in which order the activities can be executed for such cases. 

However, this already shows the central limitation of traditional process mining: the single case notion. In practice, the assumption that every event is only related to exactly one case is unrealistic.
One such example could be an event describing activity \textit{Load cargo} for objects \textit{plane1} and \textit{bag1} at \textit{2021-10-02 17:23}.
If one considers this event and the corresponding event log, what would be the case notion? The plane to be loaded with cargo or the bag to be loaded into the plane? By focusing on only one case notion one would omit the behaviour of the other case notions, e.g., focusing on only the plane would omit the process of checking in and transporting baggage. Uniting the different case notions into one single, comprehensive process model allows us to consider dependencies between the executions that are not available otherwise.

\iffalse
If one considers the ordering process at an online shop what would be the case notion? The item to be ordered, the order, the customer or the delivery? By focusing on only one case notion one would omit the behaviour of the other case notions, e.g., focusing on only the order would omit the process of picking up, packaging and delivering an item. Uniting the different case notions into one single, comprehensive process model allows us to consider dependencies between the executions that are not available otherwise. 
\fi
Object-centric process mining\cite{OCPMDivergenceAndconvergence} addresses these problems and aims to define techniques and standards for retrieving and analyzing comprehensive, easy-to-understand models of processes with multiple case notions.
An event log format \cite{OCEL} and a first discovery technique \cite{DiscoveringObjectCentricPetriNets} have already been introduced. This discovery technique yields an object-centric Petri net. However, the missing key to enable further research on object-centric process discovery on the one hand and objective model evaluation on the other hand are quality criteria that link object-centric model and log and specify the conformance of the model. In traditional process mining quality criteria like precision, fitness, simplicity and generality are used to express the quality of a model with respect to a log \cite{ProcessMiningDSIA}. These can be used to compare different models of the same event log, evaluate the results of different discovery algorithms, specify the conformance of a handmade process model and more.
\iffalse
WHAT HERE
However, process discovery is far from trivial. Compare the plethora of discovery algorithms that have been introduced over the last years \cite{ProcessMiningDSIA} to improve traditional process discovery. There are different quality measures like precision, fitness, simplicity and generality to evaluate the quality of a model with respect to a log in traditional process mining. These can be used to describe the quality of a model, compare different models with each other and assess the results of discovery algorithms to improve overall process discovery.
UNTIL HERE\fi
So far, there are no quality measures to evaluate an object-centric Petri net.
To facilitate further research we need a way to describe the quality of an object-centric Petri net, e.g., how well it fits the data or how precisely it describes the data.
Due to multiple case notions and many-to-many relationships fitness and precision notions from traditional process mining can not trivially be adapted to the object-centric setting.

In this paper, we introduce a fitness and precision notion for object-centric Petri nets with respect to object-centric event logs. Our fitness notion can be seen as an object-centric adaption to replaying traces \cite{RozinatConformanceCehcking}, the precision notion as a generalization of the escaping edges precision \cite{EscapingEdgesPrecision}. 

The paper is structured as follows. In \autoref{sec:relatedwork}, we discuss other approaches of handling multiple case notions. In \autoref{sec:ocpm}, we introduce object-centric event logs and Petri nets and illustrate them on the basis of a running example. In \autoref{sec:precision}, we formally introduce our fitness and precision notion. This is further accompanied by our running example. In \autoref{sec:algorithm}, we present an algorithm to calculate both precision and fitness. We use three different models with respect to an event log in \autoref{sec:evaluation} to evaluate our introduced notions.
We conclude this paper in \autoref{sec:conclusion}.

\section{Related Work}
\label{sec:relatedwork}
In this section, we introduce the related work on handling multiple case notions and the techniques to determine fitness and precision in traditional process mining.

The different approaches to handle multiple case notions can be grouped into three categories: interacting processes, colored Petri net approaches, and novel modeling techniques. We discuss the corresponding approaches, their disadvantages and how object-centric Petri nets relate to these.

Several approaches to handle multiple case notions use individual processes with some notion for interaction between them. The first method that was introduced is proclets \cite{ProcletsIntroduction}, describing interacting workflow processes. Over time, this concept was extended to cover many-to-many relationships \cite{ProcletManytoMany} and different analysis techniques \cite{ProcletsPerformanceSpectrum}. Another modeling technique is artifact-centric modeling  \cite{ArtifactsIntroduction,UMLbasedVerifiableIntroduction}.
Process discovery techniques for artifact-centric modeling have been developed \cite{ArtifactsDiscoveryII}, as well as conformance checking methods \cite{ArtifactsConformance,ArtifactsDecomposedConformance}. 
To deal with the complexity of artifact-centric modeling, \cite{ArtifactsWithRestrictions} introduced a restricted artifact-centric process model definition where no concurrency is allowed within one artifact.
The main disadvantage of these approaches is the absence of a single comprehensive model of the process as they show interacting individual processes. Furthermore, models tend to get  too complex  and hard to understand.

The second approach to handle multiple case notions are colored Petri nets. DB-nets \cite{DBNetsCPNsAndRelationalDatabase} introduce a modeling technique to include a data-perspective into a process model using a colored Petri net. However, the modeling is hard to understand for a user and it targeted to a modeling, not a mining setting. \iffalse \cite{RestrictedCPNsConformanceChecking,ConservativeWorkflowNetsReplayFitness} use colored Petri nets with restrictions to model a process with multiple case notions. The restrictions being put on the colored Petri net, i.e., no concurrency within one case notion and discarding variable one-to-many relationships, renders this approach infeasible in many settings.\fi     Furthermore, there is a plethora of approaches that uses colored Petri nets with some restrictions, e.g., no concurrency within one case notion and discarding variable one-to-many relationships, which restrictions render them infeasible in many settings. The discussion of these approaches is outside the scope of this paper.

The third group includes newly proposed modeling techniques to condense a process with multiple case notions to one model.
Object-Centric Behavioral Constraints (OCBC) \cite{OCBCFoundation} are a recently introduced approach to show behavior and relationships of different objects in one model. Discovery algorithms as well as quality measures have been proposed for this discovery technique \cite{OCBCDiscovery}. However, since OCBC builds on top of a data format that records the whole evolution of a database the models tend to get complex and the proposed techniques are not scalable.
Multiple View Point models \cite{StarStartMVPfromdatabase} introduce less complex models where the discovery is more scalable.  A process model can be constructed by correlating events via objects.
\cite{DiscoveringObjectCentricPetriNets} extends this approach to discover object-centric Petri nets, Petri nets with places of different colors and arcs that can consume a variable amount of tokens. Object-centric Petri nets can model a process with multiple case notions in one single model, consider one-to-many and many-to-many relations, concurrency within case notions and a scalable discovery algorithm is available. Therefore, object-centric Petri nets alleviate most of the drawbacks from other approaches in the related work with respect to multiple case notions.

Since quality metrics play an important role in traditional process mining several techniques to determine fitness and precision have been proposed. Techniques for determining fitness include causal footprints, token-based replay, alignments, behavioral recall and eigenvalue recall \cite{FitnessOverview}. In this paper, we use an adaptation of a replaying fitness, i.e., whether the preoccurring activities of an event can be replayed on the object-centric Petri net. Techniques for determining precision include escaping edges precision, negative event precision and projected conformance checking  \cite{PrecisionOverview}. The adaptation of these techniques to object-centric event logs and Petri nets each pose their own challenges. i.e., dealing with multiple case notions and variable arcs in the Petri net. In this paper, we propose a generalization of the escaping edges precision \cite{EscapingEdgesPrecision} which links an event to a state in the process model and determines the behavior allowed by model and log and, subsequently, the precision. This can also be seen as an object-centric adaptation of replaying history on the process model to determine fitness and precision \cite{ReplayingHistoryFitnessPrecision}.

\section{Object-Centric Process Mining}
\label{sec:ocpm}
\begin{figure}[t!]
    \centering
    \includegraphics[width=0.34\textwidth]{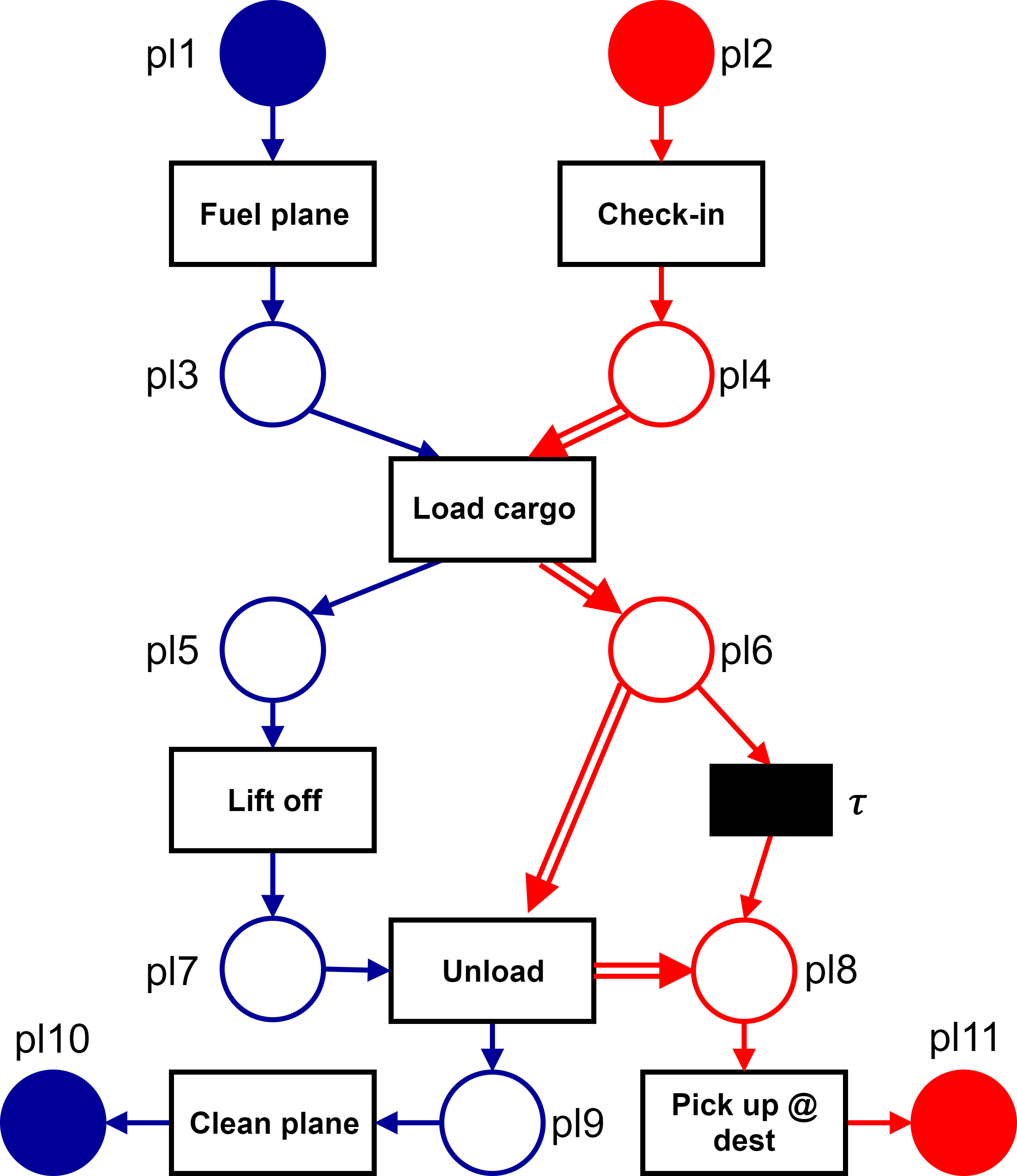}
    \caption{Example of an object-centric Petri net $\textit{OCPN}_1$.}
    \label{fig:ocpn}
\end{figure}

Object-centric process mining moves away from the single case notion of traditional process mining, i.e., every event is related to exactly one case, and opens up the possibility for one event being related to multiple, different case notions. The foundations were defined in \cite{DiscoveringObjectCentricPetriNets}.
We introduce these key concepts of object-centric process mining in this section.  These are accompanied by a running example of a flight process that considers planes and how the associated baggage is handled. It describes the operations of the plane, i.e., fueling, cleaning and lift off, and the handling of baggage, i.e., check-in, loading and pick up.
We first introduce some basic definitions and notations.
\begin{definition}[Power Set, Multiset, Sequence]
Let X denote a set. $\mathcal{P}(X)$ denotes the power set of these elements, the set of all possible subsets. A multiset $\mathcal{B}{:} X {\rightarrow} \mathbb{N}$ assigns a frequency to each element in the set and is denoted by $[x^m]$ for $x\in X$ and frequency $m\in \mathbb{N}$. A sequence of length $n$ assign positions to elements $x{\in} X$ of a set $\sigma{:} \{1,\ldots, n\} {\rightarrow} X  $. It is denoted by $\sigma {=} \langle  x_1,\ldots, x_n\rangle$. The concatenation of two sequences is denoted with $\sigma_1 {\cdot} \sigma_2$. Concatenation with an empty sequence does not alter a sequence $\sigma {\cdot} \langle\rangle {=} \sigma$. The number of elements in a sequence $\sigma$ is given by $len(\sigma)$.
\end{definition}
The central part of object-centric process mining are objects. Each object is of exactly one object type.
\begin{definition}[Object and Object Types]
 Let $\mathcal{A}$ be the universe of activities.
Let $U_o$ be the universe of objects and $U_{ot}$ be the universe of object types. The function $\pi_{otyp}:U_o \rightarrow U_{ot}$ maps each object to its type.
\end{definition}
The two universes, of objects and of object types, contain all possible objects and all possible object types. Every object $o\in U_o$ has a type $\pi_{otyp}(o)$. In our example we are considering object types $OT {=} \{\textit{planes}, \textit{baggage} \}$. We have, in total, six objects: $O {=}\{\textit{p1},\textit{p2},\textit{b1},\textit{b2},\textit{b3},\textit{b4}\}$ of which two are planes and four are baggage, indicated by the first letter of the object, i.e., $\pi_{otyp}(\textit{p1}) {=} \pi_{otyp}(\textit{p2}) {=} \textit{plane}, \pi_{otyp}(\textit{b1}) {=} \cdots {=} \pi_{otyp}(\textit{b4})\allowbreak {=} \textit{baggage}$.
In general, we are interested in recording the behaviour of objects over time, i.e., at which point in time activities were executed that concern objects. This is done using an object-centric event log.
The OCEL format \cite{OCEL} records the corresponding activity, timestamp and objects of each object type for an event. Additional attributes are stored for each object. 
We use a reduced definition of the object-centric event log to focus on the aspects relevant for this paper.
 \begin{definition}[Object-Centric Event Log]
Let $U_E$ be the universe of event identifiers. An object-centric event log is a tuple $L{=}(E,OT,O,\pi_{act},\pi_{omap}, \leq )$ of event identifiers $E {\subseteq} U_E$, object types $OT{\subseteq} U_{ot}$, objects $O{\subseteq} U_o$, a mapping function from event to activity $\pi_{act}{:} E {\rightarrow} \mathcal{A}$, a mapping function from event to related objects $\pi_{omap}{:} E {\rightarrow} \mathcal{P}(O)$ and a total order $\leq$.
 \end{definition}
This definition gives us event identifiers of which each is related to an activity and to some objects. These event identifiers are subject to a total order, i.e., by the time of their occurrence.
\iffalse
\begin{table}[]
    \centering
    \caption{Example of an Object-Centric Event Log}
    \begin{tabular}{|c|l|c|c|c|}\hline
     \multicolumn{1}{|c|}{Event ID}& \multicolumn{1}{|c|}{Activity} &\multicolumn{1}{|c|}{Timestamp}  & \multicolumn{1}{|c|}{\textit{plane}} & \multicolumn{1}{|c|}{\textit{baggage}} \\\hline
    $e_1$ & Fuel plane     & 2021-03-03 10:23 & p1 & \\\hline
    $e_2$ & Check-in & 2021-03-03 10:34 & & b1\\\hline
    $e_3$ & Check-in & 2021-03-03 10:37 & & b2\\\hline
    $e_4$ & Load cargo & 2021-03-03 11:04 & p1 & b1,b2\\\hline
    $e_5$ & Lift off & 2021-03-03 11:46 & p1 & \\\hline
    $e_6$ & Unload & 2021-03-03 15:46 & p1 & b1,b2\\\hline
    $e_7$ & Pick up @ dest & 2021-03-03 16:14 &  & b1\\\hline
    $e_8$ & Pick up @ dest & 2021-03-03 16:15 &  & b2\\\hline
    $e_9$ & Clean & 2021-03-03 16:20 & p1 & \\\hline
    $e_10$ & Fuel plane     & 2021-03-04 10:23 & p2 & \\\hline
    $e_11$ & Check-in & 2021-03-04 10:34 & & b3\\\hline
    $e_12$ & Check-in & 2021-03-04 10:37 & & b4\\\hline
    $e_13$ & Load cargo & 2021-03-04 11:04 & p2 & b3,b4\\\hline
    $e_14$ & Lift off & 2021-03-04 11:46 & p2 & \\\hline
    $e_15$ & Unload & 2021-03-04 15:46 & p2 & b3,b4\\\hline
    $e-16$ & Clean & 2021-03-04 16:12 & p2 & \\\hline
    $e_17$ & Pick up @ dest & 2021-03-04 16:14 &  & b3\\\hline
    $e_18$ & Pick up @ dest & 2021-03-04 16:15 &  & b4\\\hline
    \end{tabular}
    \label{tab:ocel_}
\end{table}
\fi
\begin{table}[t]
    \centering
    
    \caption{Example of an object-centric event log $L_1$}
    \resizebox{0.73\columnwidth}{!}{%
    \begin{tabular}{|c|l|c|c|}\hline
     \multicolumn{1}{|c|}{Event ID}& \multicolumn{1}{|c|}{Activity}   & \multicolumn{1}{|c|}{\textit{plane}} & \multicolumn{1}{|c|}{\textit{baggage}} \\\hline
    $e_1$ & \textit{Fuel plane}      & \textit{p1} & \\\hline
    $e_2$ & \textit{Check-in}  & & \textit{b1}\\\hline
    $e_3$ & \textit{Check-in}  & & \textit{b2}\\\hline
    $e_4$ & \textit{Load cargo}  & \textit{p1} & \textit{b1,b2}\\\hline
    $e_5$ & \textit{Lift off}  & \textit{p1} & \\\hline
    $e_6$ & \textit{Unload}  & \textit{p1 }& \textit{b1,b2}\\\hline
    $e_7$ & \textit{Pick up @ dest}  &  & \textit{b1}\\\hline
    $e_8$ & \textit{Pick up @ dest}  &  & \textit{b2}\\\hline
    $e_9$ & \textit{Clean}  & \textit{p1} & \\\hline
    $e_{10}$ & \textit{Fuel plane}      & \textit{p2} & \\\hline
    $e_{11}$ & \textit{Check-in}  & & \textit{b3}\\\hline
    $e_{12}$ & \textit{Check-in}  & & \textit{b4}\\\hline
    $e_{13}$ & \textit{Load cargo}  & \textit{p2} & \textit{b3,b4}\\\hline
    $e_{14}$ & \textit{Lift off}  & \textit{p2} & \\\hline
    $e_{15}$ & \textit{Unload}  & \textit{p2} & \textit{b3,b4}\\\hline
    $e_{16}$ & \textit{Clean}  & \textit{p2} & \\\hline
    $e_{17}$ & \textit{Pick up @ dest}  &  & \textit{b3}\\\hline
    $e_{18}$ & \textit{Pick up @ dest}  &  & \textit{b4}\\\hline
    \end{tabular}%
    }
    \label{tab:ocel}
\end{table}
An example of an object-centric event log is depicted in \autoref{tab:ocel}. It consists of the events with the identifiers $E{=} \{e_1,\ldots,e_{18}\}$. The event identifier is unique and provides the order of events. The activity of an event, e.g., $\pi_{act}(e_4){=}\textit{Load cargo}$ is given as well as the related objects, e.g., $\pi_{omap}(e_4){=}\{\textit{p1,b1,b2}\}$. 
Discovering process models from event data plays a central role in process mining as it is a data-driven way to uncover the true nature of a process. One way to represent process models are Petri nets\cite{ProcessMiningDSIA}. Van der Aalst and Berti \cite{DiscoveringObjectCentricPetriNets} introduce the concept of object-centric Petri nets and provide an algorithm to discover an object-centric Petri net from an object-centric event log. An object-centric Petri net is a colored Petri net where the places are colored as one of the object types and some arcs can consume a variable amount of tokens. \iffalse An object-centric Petri net consists of places and transitions. Arcs can connect a place to a transition and vice versa. Some arcs are variable arcs. Every place has exactly one object type.\fi
 \begin{definition}[Object-Centric Petri Net]
 Let $N{=}(P,T,F,l)$ be a Petri net consisting of places $P$, transitions $T$, flow relations $F{\subseteq} (T{\times} P) {\cup} (P{\times} T)$ and a labeling function $l{:}T {\nrightarrow} \mathcal{A}$.
 An object-centric Petri net is a tuple $\textit{OCPN} {=} (N,pt,F_{\textit{var}})$ of Petri net $N$, a mapping of places to object types $pt{:} P {\rightarrow} OT$ and  a set of variable arcs $F_{\textit{var}} {\subseteq} F$. We introduce the notations:
 \begin{itemize}
     \item[$\bullet$]  ${\bigcdot} t {=} \{p {\in} P \mid (p,t) {\in} F\}$ is the preset of a transition $t$.
     \item[$\bullet$]$t {\bigcdot}  {=} \{p {\in} P \mid (t,p) {\in} F\}$ is the postset of a transition $t$.
     \item[$\bullet$] $tpl(t) {=} \{pt(p) \mid p {\in} {\bigcdot} t {\cup} t {\bigcdot}\}$ are the object types associated to a transition $t$.
     \item[$\bullet$] $tpl_{nv}(t) {=} \{pt(p) {\mid} p {\in} P \wedge \{(p,t),(t,p)\} {\cap} (F{\setminus} F_{\textit{var}}) {\neq} \emptyset\}$ are the non-variable object types associated to transition $t$.
 \end{itemize} 
 \end{definition}
An example of an object-centric Petri net for the flight process is depicted in \autoref{fig:ocpn}. Places are indicated by circles and transitions are indicated by rectangles. Arcs are indicated by arrows while variable arcs are indicated by double arrows. The object type of a place is shown by its color, blue being type \textit{plane} and red being type \textit{baggage}. The label of each transition is depicted inside the transition itself. The transition with no label is a so called silent transition. We refer to this transition as $\tau$.
This definition of the Petri net itself is not sufficient to describe a process and the behavior it allows.
We need to introduce further concepts.
We, first, introduce the concept of a marking.
Places of object-centric Petri nets can hold tokens of the same object type. The marking of an object-centric Petri net can be expressed as objects being associated to places of the corresponding object type.
 \begin{definition}[Marking of an Object-Centric Petri Net]
 Let $OCPN {=} (N,pt,F_{\textit{var}})$ be an object-centric Petri net with $N{=}(P,T,F,l)$. The set of possible tokens is described by $Q_{OCPN} {=} \{(p,o) {\in} P {\times} U_o \mid pt(p) {=} \pi_{otyp}(o)\}$. A marking is a multiset of tokens $M {\in} \mathcal{B}(Q_{OCPN})$.
 \iffalse A marking is a multiset of places describing the number of tokens in each place $M\in\mathcal{B}(P)$. The marking of an object-centric Petri net is also called state. If a marking $m'$ can be reached by executing transition $t\in T$ in marking $m$ we denote this by $m\xrightarrow{t}m'$.\fi
 \end{definition}
A marking describes the current state of the Petri net. It states which objects are contained in which places.
We can move between markings by firing transitions.
The concept of a binding is used to describe this. A binding is a tuple of a transition and the involved objects of every object type. The objects of the corresponding color in the binding are consumed in the input places and produced in the output places. 
A binding is only enabled at a given marking if each input place of the corresponding transition holds at least the tokens of this object type contained in the binding.
 \begin{definition}[Binding Execution]
 Let $\textit{OCPN} {=} (N,pt,F_{\textit{var}})$ be an object-centric Petri net with $N{=}(P,T,F,l)$.
 $B {=} \{(t,b) \in T {\times} (U_{ot} {\nrightarrow} \mathcal{P}(U_o)) \mid  dom(b) {=} tpl(t) \wedge \forall_{ot {\in} tpl_{nv}(t)} |b(ot)|{=}1\}$ defines the set of all possible bindings. The consumed tokens of a binding $(t,b) {\in} B$ are defined by $cons(t,b) {=} [(p,o){\in} Q_{\textit{OCPN}} \mid p {\in} {\bigcdot} t \wedge o {\in} b(pt(p)) ]$ and the produced tokens are defined by $prod(t,b) {=} [(p,o){\in} Q_{\textit{OCPN}} \mid p {\in} t {\bigcdot}  \wedge o {\in} b(pt(p)) ]$. A binding $(t,b) {\in} B$ in marking $M$ is enabled if $cons(t,b) {\leq} M$. The execution of an enabled binding in marking $M$ leads to marking $M' {=} M {-} cons(t,b) {+} prod(t,b)$. This is denoted by $M {\xrightarrow{(t,b)}}M'$
 \end{definition}
$M{\xrightarrow{(t,b)}}M'$ indicates that the occurrence of binding $(t,b)$ in marking $M$ leads to marking $M'$. Multiple subsequent enabled bindings can be encoded as  sequence  $\sigma {=} \langle (t_1,b_1),\ldots,(t_n,b_n)\rangle {\in} B^*$. A sequence that starts in marking $M_0$ and results in marking $M_n$ is encoded as $M_0 {\xrightarrow[]{\sigma}}M_n$.
 We can take a binding at the \textit{Load cargo} transition of the Petri net in \autoref{fig:ocpn} as an example. We assume a marking where the input places of \textit{Load cargo} contain plane $p1$ and baggages $b1,b2$ according to their color $[(pl3,p1),(pl4,b1),(pl4,b2)]$. In this state of the Petri net the binding of \textit{Load cargo} and $p1,b1,b2$ is enabled since the input places of the transition contain enough tokens of objects contained in the binding. The binding can be executed and leads to a new marking where the tokens of $p1,b1,b2$ moved from the input to the output places of \textit{Load cargo}. We can construct a sequence of such enabled bindings that describes the execution of several bindings after each other. In this way, we can express that one marking is reachable from another marking in the Petri net. We define initial and final markings. Together with the object-centric Petri net these form an accepting object centric Petri net. All binding sequences starting from an initial marking and ending in a final marking form the accepted behavior of that accepting object-centric Petri net.
 \begin{definition}[Accepting Object-Centric Petri Net]
 Let $\textit{OCPN} {=}  (N ,  pt ,F_{\textit{var}})$ be an object-centric Petri net with $N{=}(P, \allowbreak T,F,l)$ and let $M_{init}, M_{final} {\subseteq} \mathcal{B}(Q_{\textit{OCPN}})$ be initial and final markings of that Petri net. $\textit{OCPN}_A {=} (\textit{OCPN},M_{init}, M_{final})$  is a tuple that describes an accepting object-centric Petri net. $\Sigma = \{\sigma {\in} B^* \mid M {\in} M_{init} \wedge M' {\in} M_{final} \wedge M_{init} {\xrightarrow[]{\sigma}} M_{final}\}$ describes all the binding sequences, i.e., the accepted behavior of the Petri net.
 \end{definition}
 In our example, an accepted behavior is a binding sequence that starts in a marking with only tokens in the places $pl1$ and $pl2$ and ends in a marking with only tokens in the places $pl10$ and $pl11$.
 The algorithm of \cite{DiscoveringObjectCentricPetriNets} can be used to discover such accepting object-centric Petri nets.
 
These concepts are sufficient to describe the behavior allowed by an object-centric Petri net. Starting in an initial marking all binding sequences are allowed that lead to a final marking. However, if we consider our event log in \autoref{tab:ocel} and our model in \autoref{fig:ocpn} the question arises of how well  the process model describes the object-centric event log.
So far research on object-centric process mining has focused on specifying standardized event log formats and discovering models. The missing key to enable further research on improving object-centric process discovery are quality criteria for a discovered object-centric model.

\section{Fitness and Precision for Object-Centric Petri Nets}
\label{sec:precision}
Generally speaking, fitness and precision compare the behaviour seen in a data set to the behaviour possible in a model. The more behavior of the data set the model allows the fitter it is. The more behaviour the model allows that is not covered by the data set the more imprecise it is. This general notion of fitness and precision can be adapted to event logs and process models by comparing the recorded activity sequences in the event log to the possible activity sequences in the process model\cite{ReplayingHistoryFitnessPrecision} in traditional process mining. \iffalse In traditional process mining the sequence of activities of the same case happening before an event is called the prefix. By taking all events that have the same prefix and collecting all activities that happened with this prefix we can describe the behaviour allowed by the event log. These prefixes can now be replayed on the model, assuming perfect fitness. All enabled activities in the state after replaying describe the behaviour allowed by the process model. The precision can be calculated by doing this for each prefix in the event log. This is one possible way of determining precision for traditional event logs, called escaping edges precision.\fi However, this can not easily be adapted to object-centric event logs and Petri nets.
The main problem is multiple case notions of an event. 
There is no single sequence of previously occurring activities for an event anymore. Multiple involved objects are associated to different activity sequences that were previously executed for them. Furthermore, even if there is only one object involved in an event, an activity sequence of one object might be dependent on the occurrence of an activity sequence of another object. Compare the exemplary event log in \autoref{tab:ocel}. Before event $e_5$, \textit{Lift off} of plane $p1$, happens, the baggages $b1$ and $b2$ have to be loaded into the plane, making this event also dependent on these two objects even though they are not included in the event itself.
We define a context notion that describes the previously executed activity sequences of the objects on which an event depends. The dependent objects and their relevant events are extracted by constructing the event-object graph.
\begin{definition}[Event-Object Graph and Context of an Event]
\label{def:context}
Let $L{=}(E,OT,O,\pi_{act},\pi_{omap}, \leq )$ be an object-centric event log, let $ot {\in} OT$ be an object type and let $o {\in} O$ be an object.
We introduce the following notations:
\begin{itemize}
    \item[$\bullet$] $G_L{=}(V,K)$ is the event-object graph of the event log $L$ with $V{=}E$ and $K{=}\{(e',e) \in E {\times} E \mid e' {<} e \wedge \pi_{omap}(e) \cap \pi_{omap}(e') \neq \emptyset \}$.
     \item[$\bullet$] $\circ e {=} \{e'{\in} V \mid \exists \sigma {=} \langle e',\ldots, e \rangle {\in} E^* \; \forall i {\in} \{ 1,\ldots,len(\sigma){-}1\}\allowbreak (\sigma_i,\sigma_{i+1}) \in K\}$ is the event preset of an event $e\in E$.
    \item[$\bullet$] $\circ e \downarrow_o {=} \langle \pi_{act}(e_1), \ldots, \pi_{act}(e_n)  \rangle$ such that \begin{itemize}
    \item[$\bullet$] $\{e'{\in} \circ e \mid  o {\in} \pi_{omap}(e')\} {=} \{e_1,\ldots,e_n\}$
    \item[$\bullet$] $e_1 < \ldots < e_n$
    \end{itemize}
    \iffalse
    \item[$\bullet$]  $\textit{prefix}_e(o) = \langle \pi_{act}(e_1), \ldots, \pi_{act}(e_n), \pi_{act}(e)  \rangle$ such that \begin{itemize}
    \item[$\bullet$]  $\{e' \in E \mid  e'< e \wedge o \in \pi_{omap}(e)\} = \{e_1, \ldots , e_n\}$
    \item[$\bullet$]  $e_1 < \ldots < e_n < e $
\end{itemize}
\fi
\end{itemize}
The context of an event is defined as $\textit{context}_e(ot) = [{\circ} e {\downarrow_o} \mid o {\in} \bigcup_{e' {\in} {\circ} e {\cup} \{e\}} \pi_{omap}(e') \wedge \pi_{otyp}(o) {=} ot]$ for any $ot {\in} OT$.\iffalse \footnote{For objects that are only associated to the event but none of the events in the event preset this will yield the empty sequence.}\fi
\iffalse
The context of an event is defined as $\textit{context}_e(ot) = [\textit{prefix}_{e'}(o) \mid e' \in \circ e \wedge o \in \pi_{omap}(e') \wedge \neg \exists e'' \in \circ e ( e'<e'' \wedge o \in \pi_{omap}(e'') ) \wedge \pi_{otyp}(o) = ot]$ for any $ot \in OT$.
\fi
\end{definition}
\begin{figure}
    \centering
    \includegraphics[width=0.6\linewidth]{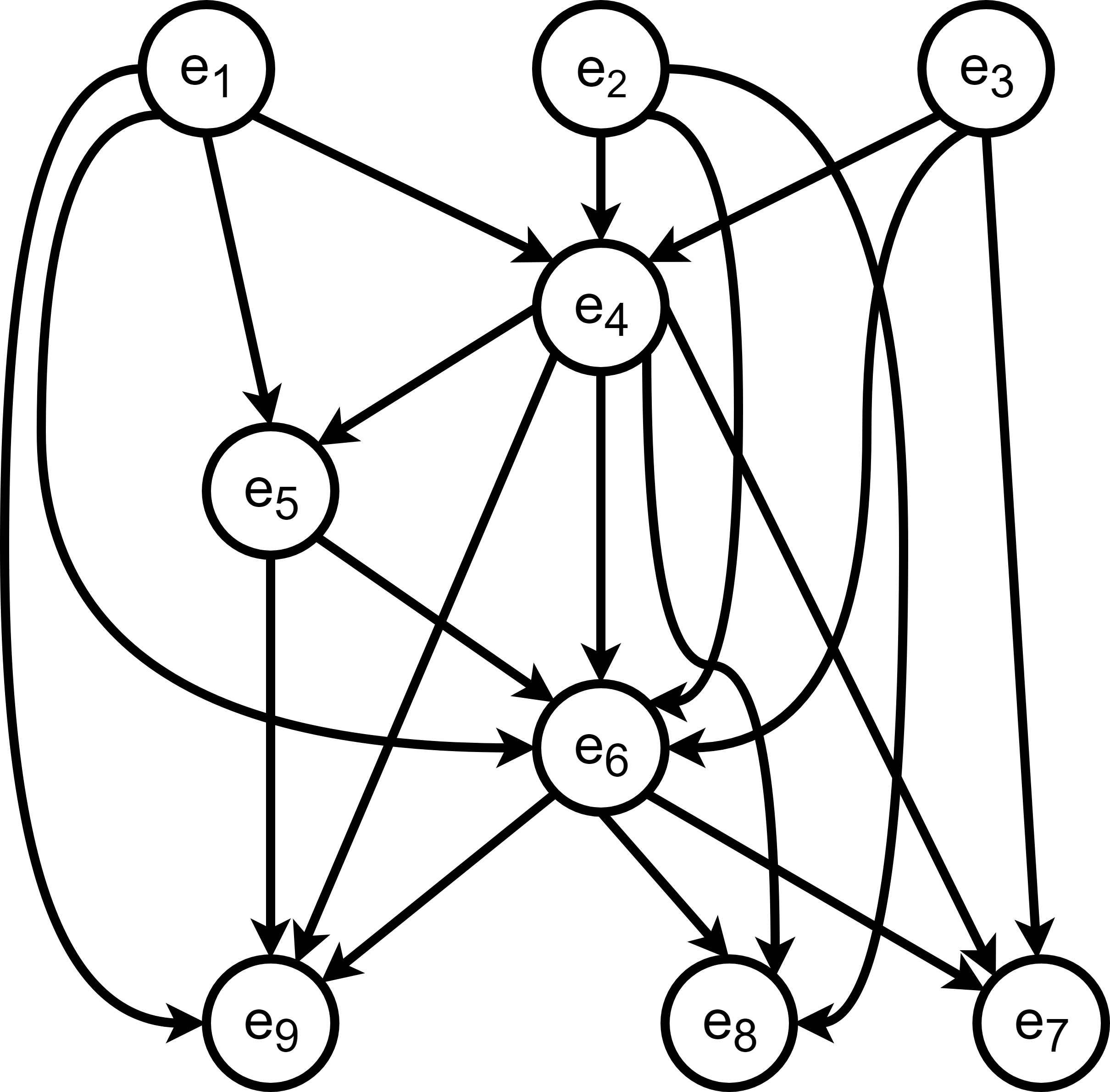}
    \caption{Event-object graph of events $\{e_1,\ldots, e_9\}$ of log $L_1$}
    \label{fig:event_object_graph}
\end{figure}
Analogously to the single case notion, each single object has a simple sequence of its previously occurring activities. We use a multiset of sequences to cover the possibility of an event being related to multiple objects of the same object type with the same sequence. We use the event-object graph to grasp every object on which the execution of an event depends and up to which event it depends on this object. The event-object graph introduces a directed edge between events if they share an object. For an event $e$, all other events that have a directed path to event $e$ form the event preset of $e$, the events on which the execution of $e$ depends. The full event-object graph of events $e_1$ to $e_9$ of log $L_1$ is depicted in \autoref{fig:event_object_graph}. Take event $e_5$ as an example. The event preset of $e_5$ is formed by all events that are directly or transitively connected to $e_5$, i.e., that have a directed path the $e_5$ in the event-object graph. This includes $e_1$ and $e_4$ but also $e_2$ and $e_3$ even though they do not share any objects with $e_5$. Therefore, $\circ e_5 = \{e_1,e_2,e_3,e_4\}$. This shows that the event preset includes all events on which the execution of an event depends, also the transitive dependencies. To construct the context we take all objects that appear in the event preset and the event itself, map them onto their sequence of activities in the event preset and construct a multiset of the objects' sequences for each object type. For $\circ e_5$ this includes the objects $p1,b1,b2$. 
\iffalse
The object history is used to grasp every object in every event that happened before the considered event and is linked to the occurrence of this event. By defining it recursively we are able to retrieve such objects that are not included in the considered event but are still relevant. When constructing the context of the object history we consider only the last event of any object to construct the prefix.
We illustrate the construction of the context with an example from \autoref{tab:ocel}. Consider the event with ID $5$. The object history of this event is $\{(p1,4),(p1,1),(b1,4),(b1,2),(b2,4),(b2,3)\}$. Only the object event pairs are considered where there is no other pair with the same object and a later event. In our example this would be $(p1,4), (b1,4)$ and $(b2,4)$.\fi We construct the sequences of these and combine them to the context $\textit{context}_{e_5}( \textit{plane}){=}\allowbreak[\langle \textit{Fuel plane}, \allowbreak \textit{Load cargo}\rangle], \allowbreak \textit{context}_{e_5}(\textit{baggage}){=} \allowbreak [\langle \textit{Check-in}, \allowbreak \textit{Load cargo}\rangle^2]$

The context can colloquially be described as everything that had to happen for this event to occur. We use the context to link log and model behaviour. The context specifies several sequences of activities executed for different object types. A binding sequence of an object-centric Petri net, thus, also has a context if we consider the executed activity sequences of all the objects in this binding sequence.
\begin{definition}[Context of a Binding Sequence]
Let $\textit{OCPN}_A {=}  (\textit{OCPN},M_{init},M_{final})$ be an accepting object-centric Petri net with $\textit{OCPN} {=} (N, \allowbreak  pt ,F_{\textit{var}})$ and $N{=} (P, T,F,l) $. Let $\sigma {=} \langle (t_1,b_1), \ldots, (t_n,b_n) \rangle$ be a sequence of enabled bindings of that Petri net. We introduce the following notations:
\begin{itemize}
    \item[$\bullet$] $(t,b){\downarrow}_o {=}  \langle l(t)\rangle $ if $ o {\in} b(\pi_{typ}(o)) {\wedge} t{\in} dom(l)$ otherwise $ (t,b){\downarrow}_o {=} \langle \rangle $.
    \item[$\bullet$] $\sigma{\downarrow}_o {=}  (t_1,b_1){\downarrow}_o {\cdot} \ldots {\cdot}  (t_n,b_n){\downarrow}_o  $ is the prefix of an object for a binding sequence.
    \item[$\bullet$] $\textit{context}_\sigma(ot) {=} [\sigma {\downarrow}_o {\mid} o {\in} \bigcup_{i {\in} \{1,\ldots,n\}} b_i(ot)] $ for any $ ot {\in} U_{ot}$ is the multiset of prefixes in a binding sequence, also called context of the binding sequence.
\end{itemize}
\end{definition}
Given a binding sequence, we can project it onto the transition labels for each object that is included in at least one of the bindings. We do this by projecting each binding of the sequence onto its transition label if the object is contained in the binding. If the transition is a silent transition, i.e., it has no label, this projection yields the empty sequence. The labels are concatenated into a sequence of labels. This is the prefix for an object. The prefixes are united into a multiset for each object type. To illustrate this we use the Petri net from \autoref{fig:ocpn} and a binding sequence. In our example, a binding consists of a transition name and bounded objects $b$ of each object type: $\allowbreak \sigma_1 {=} \allowbreak\langle (\textit{Fuel \allowbreak Plane}, \allowbreak b(\textit{plane})=\{\textit{p1}\}, \allowbreak b(\textit{baggage})=\{\}), \allowbreak (\textit{Check\allowbreak-in}, \allowbreak b(\textit{plane})=\{\}, \allowbreak b(\textit{baggage})=\{\textit{b1}\}),\allowbreak (\textit{Check\allowbreak-in}, \allowbreak b(\textit{plane})=\{\}, \allowbreak b(\textit{baggage})=\{\textit{b2}\}),\allowbreak (\textit{Load \allowbreak cargo}, \allowbreak b(\textit{plane})=\{\textit{p1}\}, \allowbreak b(\textit{baggage}){=}\{\textit{b1,b2}\})\rangle$. All objects that appear in this sequence are $\{\textit{p1,b1,b2}\}$. The projected prefix for \textit{p1} is $\langle \textit{Fuel plane},\textit{Load cargo}\rangle $. For each \textit{b1} and \textit{b2} this sequence is $\langle \textit{Check-in},\allowbreak \textit{Load cargo}\rangle $.
The context for this binding sequence is, therefore, $\textit{context}_{\sigma_1} {=} \textit{context}_{e_5}$.

Given a certain context one can look at all the possible binding  sequences that have this context. All of the states that are reached after executing any of these binding sequences are the states that are reachable given this context.
\begin{definition}[Context Reachable States]
Let $L=(E,OT,O,\pi_{act},\pi_{omap}, \leq )$ be an object-centric event log, $\textit{OCPN}_A$ be an accepting object-centric Petri net and $e {\in} E$ be an event. 
We assume the existence of an oracle $\textit{states}{:} (\textit{OCPN}_A,\textit{context}_e){\rightarrow} \mathcal{P}(M)$ that retrieves the reachable states with a binding sequence of  $\textit{context}_e$.\iffalse
All reachable states of the Petri net for the context are defined by $\textit{states}_{\textit{OCPN}}(\textit{context}_e) = \{M \in \mathcal{B}(Q_{\textit{OCPN}}) \mid \sigma \in B^* \wedge M_{init} \xrightarrow[]{\sigma} M \wedge \textit{context}_\sigma = \textit{context}_e \}$.\fi
\end{definition}
For a given context one can collect all the states that are reachable from the initial marking by a binding sequence that produces the same context. We illustrate this with the model of \autoref{fig:ocpn} and $\textit{context}_{e_5}$. The binding sequence $\sigma_1$ has the same context and results in marking $[(\textit{p5,p1}),(\textit{p6,b1}),(\textit{p6,b2})]$. There are four more binding sequences: $\sigma_2 {=} \sigma_1 {\cdot} (\tau,\{\textit{b1}\})$, $\sigma_3 {=} \sigma_1 {\cdot} (\tau,\{\textit{b2}\})$, $\sigma_4 {=} \sigma_1 {\cdot} (\tau,\{\textit{b1}\}){\cdot} (\tau,\{\textit{b2}\})$ and $\sigma_5 {=} \sigma_1 {\cdot}\allowbreak  (\tau,\{\textit{b2}\}){\cdot}\allowbreak (\tau,\{\textit{b1}\})$.\iffalse \footnote{We omit the notation for mapping of object type to objects in the binding here since it is clear to which object types the objects belong.}\fi  These are binding sequences that also execute the silent transition. As this does not influence the context they all have the same context. All these possible binding sequences define the four reachable states of the context: $\textit{states}(\textit{OCPN}_1,\textit{context}_{e_5})=\allowbreak \{[(\textit{pl5,p1}), \allowbreak (\textit{pl6,b1}), (\textit{pl6,b2})],$ $[(\textit{pl5,p1}), (\textit{pl8,b1}), (\textit{pl6,b2})], $ $ [(\textit{pl5,p1}), (\textit{pl6,b1}),\allowbreak (\textit{pl8,b2})], [(\textit{pl5,p1}), (\textit{pl8,b1}),\allowbreak (\textit{pl8,} \allowbreak \textit{b2})]\}$.

The allowed behaviour of the model is specified by the enabled activities of the model in any of the reachable states of a context.
\begin{definition}[Enabled Model Activities]
\label{def:enabled_model_activities}
Let $L=(E,OT,O,\pi_{act},\pi_{omap}, \leq )$ be an object-centric event log, $e{\in} E$ be an event, $\textit{OCPN}_A {=}  (\textit{OCPN},M_{init},M_{final})$ be an accepting object-centric Petri net with $\textit{OCPN} {=} (N, \allowbreak  pt ,F_{\textit{var}})$ and $N{=} (P, T,F,l) $. $en_{\textit{OCPN}}(e) {=} \{ l(t) {\mid} (t,b) {\in} B \wedge  \exists M {\in} \textit{states}(\textit{OCPN}_A,\textit{context}_e) \exists M' {\in} \mathcal{B}(Q_{\textit{OCPN}}) M \xrightarrow{(t,b)} M'\}$ describes the enabled activities in the model for the corresponding context of an event.
\end{definition}
Applying this to our running example, we extract the enabled activities for any state in $\textit{states}(\textit{OCPN}_1,\textit{context}_{e_5})$.  There are the two enabled activities \textit{Lift off} and \textit{Pick up} \iffalse. This means, given that two pieces of baggage were checked in and loaded into a plane that was fueled before, the model states that the two actions which can happen afterwards are the plane lifting off or the baggage being picked up at the destination. The enabled model activities are \fi, i.e., $en_{\textit{OCPN}_1}(e_5) = \{\textit{Lift off}, \textit{Pick up @ dest}\}$.

With the so far introduced concepts we can already derive the context of an event and state the possible behavior of the model for this context. To retrieve precision and fitness of the model we now need to specify the behavior that is given by the log.
The behavior recorded in the event log is specified by comparing the subsequent activities for the same context.
\begin{definition}[Enabled Log Activities]
\label{def:enabled_log}
Let  $L = \allowbreak (E,OT,O,\pi_{act},\pi_{omap}, \leq )$ be an object-centric event log and $e {\in} E$ be an event. $en_{\text{L}}(e){=} \{\pi_{act}(e') \mid e' {\in} E \wedge \textit{context}_{e'} {=} \textit{context}_e\}$ defines the enabled log activities for the corresponing context of an event.
\end{definition}
We illustrate that using our running example of $\textit{context}_{e_5}$. There is one other event that has the same context which is $e_{14}$. The activity that is executed for both events of this context is \textit{Lift off}, i.e., $en_{L_1}(e_5)=\{\textit{Lift off}\}$. \iffalse This means, given that two pieces of baggage were checked in and loaded into a plane which was fueled before, the log states that the next thing to happen is the plane lifting off and only that.\fi

The enabled log and model activities are calculated for the context of each event and the share of behaviour contained in the log and also allowed by the model, the fitness, is calculated. If all the behavior of the log is also allowed in the model it has a fitness of 1, if all replaying ends up in a final marking one could speak of perfect fitness.

\begin{definition}[Fitness]
Let $L{=}(E,OT,O,\pi_{act},\pi_{omap}, \leq )$ be an object-centric event log and $\textit{OCPN}_A$ be an accepting object-centric Petri net. The fitness of $\textit{OCPN}_A$ with respect to L is $\textit{fitness}(L,\textit{OCPN}_A) = \frac{1}{\mid E \mid}\sum_{e\in E} \frac{|en_L(e)\cap en_{\textit{OCPN}_A}(e)|}{|en_{L}(e)|}$.
\end{definition}

The enabled log and model activities are calculated for the context of each event and the share of behaviour allowed by the model and also contained in the log, the precision, is calculated. Not replayable events are skipped. If all the behavior allowed by the model is also contained in the log the model is perfectly precise. 
\begin{definition}[Precision]
Let $L{=}(E,OT,O,\pi_{act},\pi_{omap}, \leq )$ be an object-centric event log and $\textit{OCPN}_A$ be an accepting object-centric Petri net. $E_f {=} \{e'{\in} E \mid en_{\textit{OCPN}_A}(e') {\neq} \emptyset\}$ is the set of replayable events. The precision of $\textit{OCPN}_A$ with respect to L is calculated by $\textit{precision}(L,\textit{OCPN}_A) = \frac{1}{| E_f |}\sum_{e\in E_f} \frac{|en_L(e)\cap en_{\textit{OCPN}_A}(e)|}{|en_{\textit{OCPN}_A}(e)|}$.
\end{definition}
The fitness and precision metrics retrieve single comprehensive numbers about the quality of the model. We apply this to our running example. For all the events the enabled activities of the log are also allowed by the model. The fitness of the model is, therefore, $\textit{fitness}(L_1,\textit{OCPN}_1) {=}1$. The only events where the enabled model activities exceed the enabled log activities are events $e_5,e_6,e_{14},e_{15}$. For each of these events, \textit{Pick up @ dest} is enabled in the model but not in the log, i.e., the model allows for baggage to be picked up before the baggage was unloaded which is not contained in the event log. We, therefore, retrieve a precision of $\textit{precision}(L_1,\textit{OCPN}_1) {=} 0.89$. 
\section{Calculating Precision and Fitness}
\label{sec:algorithm}
\begin{figure}[t]
\removelatexerror
\begin{algorithm*}[H]
 \textbf{Input} Event log and context\;
\textbf{Output} Enabled model activities of this context\;
 collect all events that have this context\;
 \For{each event with this context}{
 extract the binding sequence of visible transitions from the event preset\;
 create initial state and append to queue\;
 \While{state in queue}{
 pop first state of the queue\;
 \If{binding sequence is fully replayed in this state}{Add enabled activities of this state to the results\;}
 \eIf{next binding enabled}{execute binding and enqueue the resulting  state\;}{enqueue all resulting states from enabled bindings of silent transitions\;}
 }
 }
 \caption{Enabled model activities of a context}
 \label{alg:calculation}
\end{algorithm*}
\end{figure}
\begin{figure*}
\begin{tabular}[t]{cc}
     a) \includegraphics[width=0.5\linewidth]{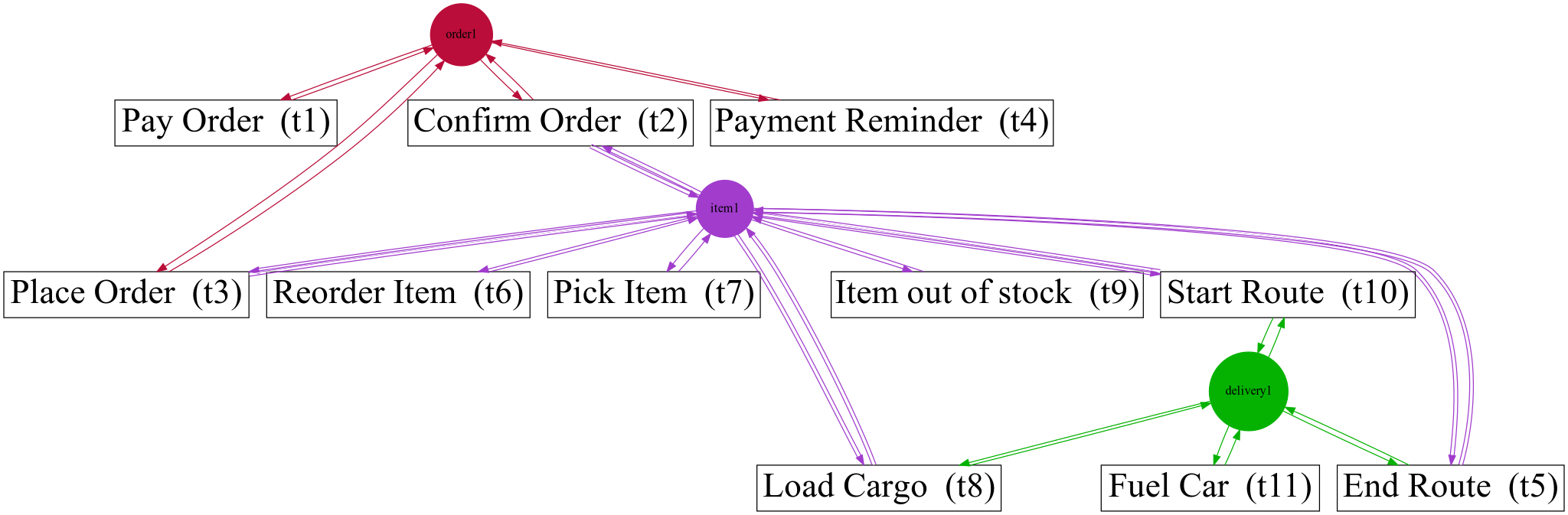}& c) \multirow[c]{2}{*}{\includegraphics[width=0.3\linewidth]{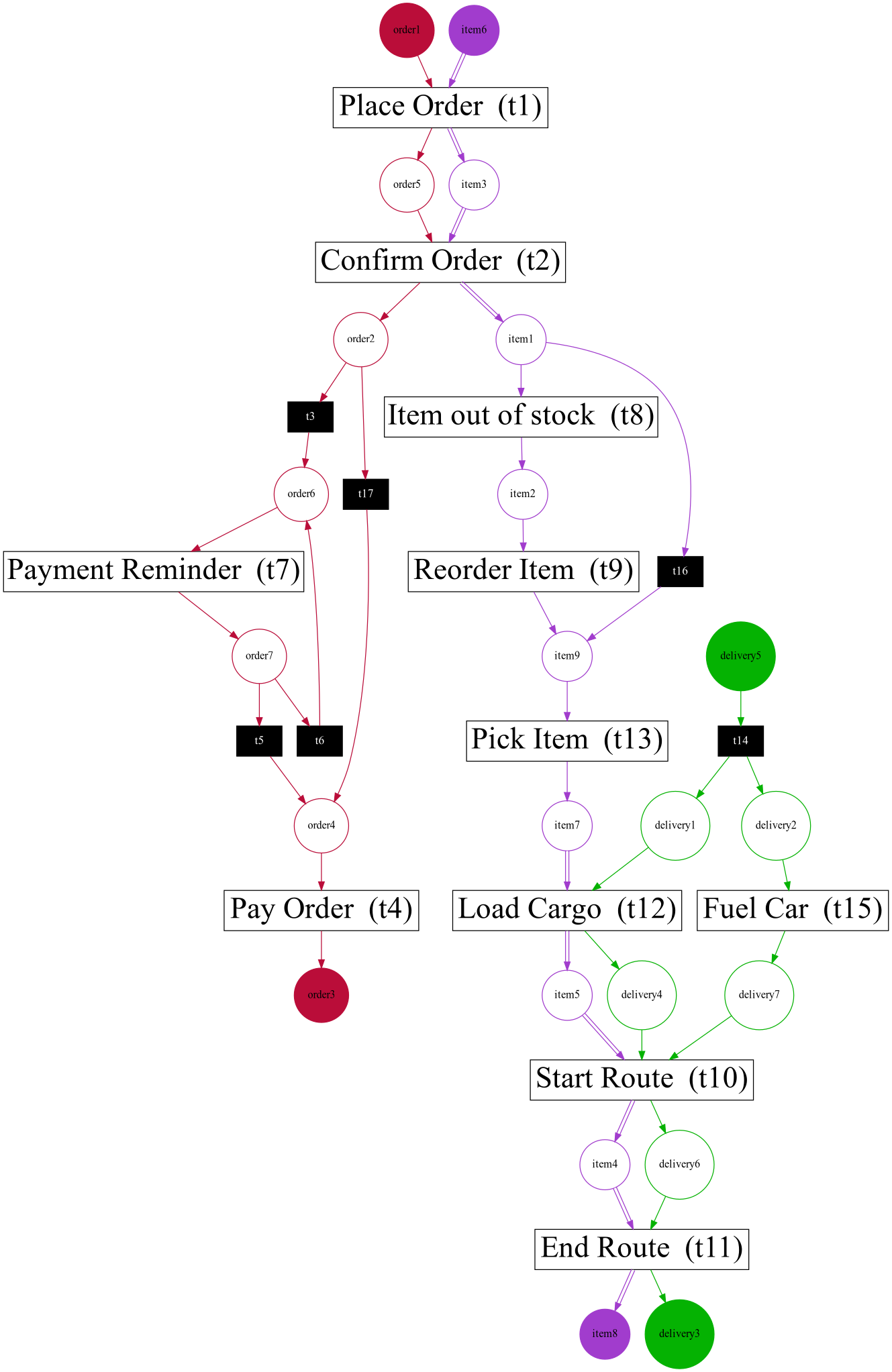}}  \\ b)
     \includegraphics[width=0.27\linewidth]{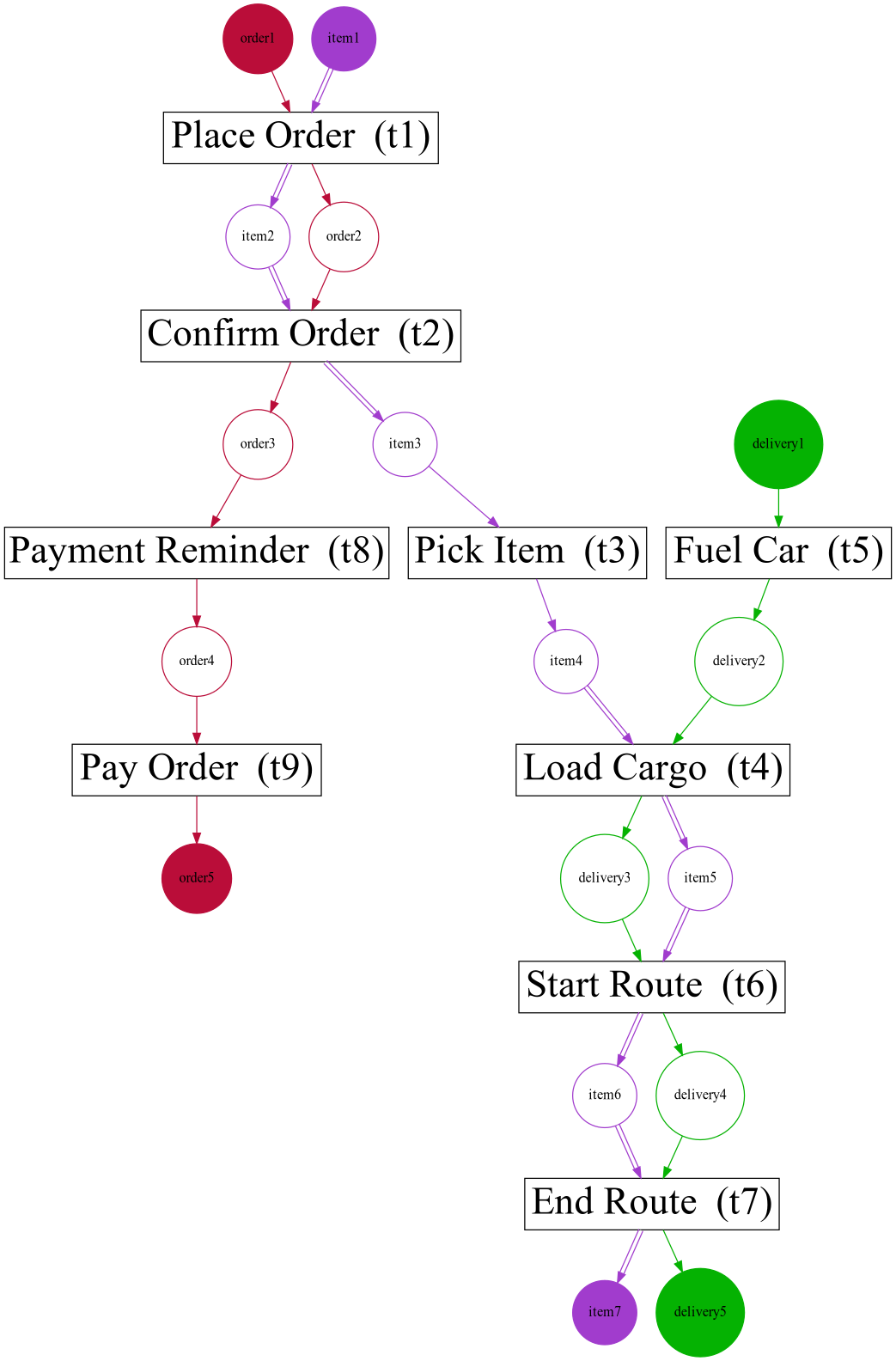} 
\end{tabular}
\caption{a) Object-centric flower model, b) restricted model, c) appropriate model}
\label{fig:models}
\end{figure*}
In this section, we discuss an algorithm to calculate precision and fitness. Our algorithm is based on replaying the events on the model. The implementation of our algorithm is available on GitHub\footnote{https://github.com/niklasadams/PrecisionFitnessOCPM}. 
The calculation of fitness and precision can be divided in two steps: determining the enabled log activities and the enabled model activities.

Constructing the contexts and calculating the enabled log activities is straightforward according to \autoref{def:context} and \autoref{def:enabled_log}. We construct the event-object graph for the event log, extract the event preset for each event, calculate the prefix for each object and merge these prefixes into the context. We, subsequently, collect the activities of all events with the same context as the enabled activities of this context.

Due to variable arcs and silent transitions the calculation of enabled model activities is not trivial. We, therefore, introduce an algorithm to determine the enabled model activities which is depicted in \autoref{alg:calculation}.
The core idea of our algorithm is to replay the events for each context and determine the enabled activities in the resulting states of the object-centric Petri net.
For a given context we collect all the events that have this context. For each of these events we extract the binding sequence of visible transitions recorded in the event log.
We do this by projecting the event preset of an event onto the bindings described by the events, i.e., transition and objects. 
See the event log $L_1$ in \autoref{tab:ocel} and event $e_5$ with event preset $\circ e_5 {=} \{e_1,e_2,e_3,e_4\}$ as example. The corresponding binding sequence of visible transitions that need to be replayed are the bindings of $e_1$ to $e_4$, consisting of the transition and recorded objects, in order of their occurrence.
We replay each binding sequence under consideration of silent transitions. This is a breadth-first search through the states reachable by executing the binding sequence and silent transitions. When the binding sequence is fully replayed all enabled activities are added to the set of enabled model activities for this context.

\iffalse

Our algorithm is an approximation for the \textit{states} oracle proposed in \autoref{def:enabled_model_activities}. As we only replay the binding sequences defined by the event log there might be binding sequences with the same context that are not considered. These are binding sequences where binding cardinalities are chosen differently for many-to-many relationships. In exchange, this approximation does not suffer from a state space explosion that one would suffer from when replaying all possible cardinalities for many-to-many relationships with many objects involved. We see it as a reasonable trade-off for applicability in practice.
\fi
%instead of discussing vague notion of approximation, disucssing the running itme
One important aspect for the practicality of these fitness and precision notion is the complexity of the algorithm. The computation of the measures consists of computing enabled log and enabled model activities. Determining enabled log activities is done in quadratic time since the event log has to be traversed once for each event to determine the preset. The computation of enabled model activities depends on the log and the Petri net.  It depends linearly on the number of events, however, two factors can lead to an exponential increase in the computation time: silent transitions and the number of objects. Coherent clusters of silent transitions with choices within one object type lead to the necessity to replay through all the possible reachable states to align the process model with the log, if possible. Especially when considering that multiple objects of one object type can be involved the state space that needs to be searched grows exponentially. This is a problem when considering, e.g., object-centric Petri nets mined by the inductive miner on a log with, e.g., noise.

\section{Evaluation}
\label{sec:evaluation}
In this section, we discuss our precision and fitness notion by applying it to three different models of one synthetic event log and analyzing the results.
We do this to assess whether our precision and fitness notion can be interpreted analogously to the notions of traditional process mining, i.e., provide an intuitive interpretation for experts and practitioners.
We use an object-centric flower model, a model that is tailored to the most frequent process execution and an appropriate model. 

\iffalse
\begin{figure}[t]
    \centering
    \includegraphics[width=1.\linewidth]{pictures/flower.png}
    \caption{Object-centric flower model}
    \label{fig:flower}
\end{figure}\fi
The first model is an object-centric adaptation to a flower model, it is depicted in \autoref{fig:models}a. In traditional process mining, a flower model is used to describe a process where every transition can be executed at any time. The fitness is very high since it covers any behavior seen in the log. The precision, however, is very low since all transitions can happen at any time, also behavior that is not covered by the event log. For an object-centric flow model, we expect a high fitness and low precision.

The second model is a model that just accounts for the most frequent activity sequence of each object type, it is depicted in \autoref{fig:models}b. This is a very restrictive model that only allows for little behavior. In traditional process mining, the precision of such a model is high as the little behavior it covers is contained in the log. The fitness, however, would be low since such a model only suits the most frequent execution of the process. We expect a low fitness and high precision for our notions.

The third model is an appropriate representation of the underlying process, it is depicted in \autoref{fig:models}c. We, therefore, expect high measures for fitness and precision.

\iffalse The first model describes the process well, we call it the \enquote{appropriate model}. It is shown in \autoref{fig:goodmodel}.  The second model is an object-centric flower model shown in \autoref{fig:flower}. The places are connected to each of the object types' activities such that each activity can be executed at any time. The third model describes only the most frequent execution of the process, we call it \enquote{restricted model}. It is shown in \autoref{fig:underfit}. \fi
\iffalse
\begin{figure}[t]
    \centering
    \includegraphics[width=.60\linewidth]{pictures/underfit.png}
    \caption{Model that is restricted to the most frequent execution of the process}
    \label{fig:underfit}
\end{figure}\fi
The results calculated by our algorithm are displayed in \autoref{tab:results}. The appropriate model fits the event log perfectly and has a precision of 0.57\footnote{On the first look this precision looks low for an appropriate model. However, consider that the object-centric Petri net makes no limitation on the order of activity executions between object types. If \textit{Pay Order} is always executed before \textit{End Route} this is not represented by the Petri net and, therefore, an imprecision.}. This is a high precision compared to the flower model which also fits perfectly but has a low precision since it always allows for every activity to be executed. The restricted model shows an almost perfect precision. However, missing concurrency, choice and activities lead to a very low fitness and approximately 54\% of the events where the context can not be replayed and which can not be considered. In summary, our fitness and precision notion behave analogously to the ones of traditional process mining and, therefore, yield an intuitive and easy-to-understand object-centric definition of fitness and precision.
\begin{table}[t]
    \centering
    \caption{Results for different models}
    \resizebox{0.99\columnwidth}{!}{%
    \begin{tabular}{|l|c|c|c|}\hline
         & Fitness & Precision & Skipped Events \\\hline
        Flower Model (\autoref{fig:models}a) &$1$&$0.25$&$0\%$ \\\hline
        Restricted Model (\autoref{fig:models}b) &$0.31$&$0.95$&$54\%$ \\\hline
        Appropriate Model (\autoref{fig:models}c) & $1$ & $0.57$& $0\%$\\\hline
    \end{tabular}
    }
    \label{tab:results}
\end{table}

\section{Conclusion}
\label{sec:conclusion}
In this paper, we introduced a precision and fitness notion for object-centric Petri nets with respect to an object-centric event log. We use the concept of a context to relate log and model behavior and calculate the enabled activities for the context in the model and the log. \iffalse By taking the share of enabled activities that is enabled in the model and also in the log we can calculate the precision for an object-centric Petri net.\fi We handle contexts that are not replayable on the Petri net by excluding them from the precision calculation. We provide an algorithm and implementation of calculating these quality metrics. We evaluate our contributions by comparing the quality measures for one event log and three different models. Our fitness and precision notion offer an objective way to evaluate the quality of an object-centric Petri net with respect to an object-centric event log. In the future, we want to use these concepts for evaluation of improved object-centric process discovery. Other future lines of research could focus on handling non-replayable context, e.g., by the calculation of object-centric alignments or on providing approximations for models with large state spaces for replay. Furthermore, other quality metrics considered in traditional process mining, i.e., generality and simplicity, could be investigated.

\bibliographystyle{IEEEtran}
\bibliography{bibliography}

% Generated by IEEEtran.bst, version: 1.14 (2015/08/26)
\begin{thebibliography}{10}
\providecommand{\url}[1]{#1}
\csname url@samestyle\endcsname
\providecommand{\newblock}{\relax}
\providecommand{\bibinfo}[2]{#2}
\providecommand{\BIBentrySTDinterwordspacing}{\spaceskip=0pt\relax}
\providecommand{\BIBentryALTinterwordstretchfactor}{4}
\providecommand{\BIBentryALTinterwordspacing}{\spaceskip=\fontdimen2\font plus
\BIBentryALTinterwordstretchfactor\fontdimen3\font minus
  \fontdimen4\font\relax}
\providecommand{\BIBforeignlanguage}[2]{{%
\expandafter\ifx\csname l@#1\endcsname\relax
\typeout{** WARNING: IEEEtran.bst: No hyphenation pattern has been}%
\typeout{** loaded for the language `#1'. Using the pattern for}%
\typeout{** the default language instead.}%
\else
\language=\csname l@#1\endcsname
\fi
#2}}
\providecommand{\BIBdecl}{\relax}
\BIBdecl

\bibitem{ProcessMiningDSIA}
W.~M.~P. van~der Aalst, \emph{Process mining: Data science in action}.\hskip
  1em plus 0.5em minus 0.4em\relax Springer, 2016.

\bibitem{OCPMDivergenceAndconvergence}
------, ``Object-centric process mining: Dealing with divergence and
  convergence in event data,'' in \emph{{SEFM} 2019.}, ser. LNCS, vol.
  11724.\hskip 1em plus 0.5em minus 0.4em\relax Springer, 2019, pp. 3--25.

\bibitem{OCEL}
A.~F. Ghahfarokhi, G.~Park, A.~Berti, and W.~M.~P. van~der Aalst, ``{OCEL:} {A}
  standard for object-centric event logs,'' in \emph{{ADBIS} (Short Papers)
  2021}, ser. CCIS, vol. 1450.\hskip 1em plus 0.5em minus 0.4em\relax Springer,
  pp. 169--175.

\bibitem{DiscoveringObjectCentricPetriNets}
W.~M.~P. van~der Aalst and A.~Berti, ``Discovering object-centric {P}etri
  nets,'' \emph{Fundam. Informaticae}, vol. 175, no. 1-4, pp. 1--40, 2020.

\bibitem{RozinatConformanceCehcking}
A.~Rozinat and W.~M.~P. van~der Aalst, ``Conformance checking of processes
  based on monitoring real behavior,'' \emph{Inf. Syst.}, vol.~33, no.~1, pp.
  64--95, 2008.

\bibitem{EscapingEdgesPrecision}
J.~Munoz{-}Gama and J.~Carmona, ``A fresh look at precision in process
  conformance,'' in \emph{{BPM} 2010.}, ser. LNCS, vol. 6336.\hskip 1em plus
  0.5em minus 0.4em\relax Springer, pp. 211--226.

\bibitem{ProcletsIntroduction}
W.~M.~P. van~der Aalst, P.~Barthelmess, C.~A. Ellis, and J.~Wainer, ``Workflow
  modeling using proclets,'' in \emph{CoopIS 2000.}, ser. LNCS, vol.
  1901.\hskip 1em plus 0.5em minus 0.4em\relax Springer, pp. 198--209.

\bibitem{ProcletManytoMany}
D.~Fahland, ``Describing behavior of processes with many-to-many
  interactions,'' in \emph{{PETRI} {NETS} 2019.}, ser. LNCS, vol. 11522.\hskip
  1em plus 0.5em minus 0.4em\relax Springer, pp. 3--24.

\bibitem{ProcletsPerformanceSpectrum}
V.~Denisov, D.~Fahland, and W.~M.~P. van~der Aalst, ``Multi-dimensional
  performance analysis and monitoring using integrated performance spectra,''
  in \emph{{ICPM} 2020}, ser. {CEUR}, vol. 2703.\hskip 1em plus 0.5em minus
  0.4em\relax CEUR-WS.org, pp. 27--30.

\bibitem{ArtifactsIntroduction}
D.~Cohn and R.~Hull, ``Business artifacts: {A} data-centric approach to
  modeling business operations and processes,'' \emph{{IEEE} Data Eng. Bull.},
  vol.~32, no.~3, pp. 3--9, 2009.

\bibitem{UMLbasedVerifiableIntroduction}
D.~Calvanese, M.~Montali, M.~Esta{\~{n}}ol, and E.~Teniente, ``Verifiable {UML}
  artifact-centric business process models,'' in \emph{{CIKM} 2014.}\hskip 1em
  plus 0.5em minus 0.4em\relax {ACM}, pp. 1289--1298.

\bibitem{ArtifactsDiscoveryII}
X.~Lu, M.~Nagelkerke, D.~van~de Wiel, and D.~Fahland, ``Discovering interacting
  artifacts from {ERP} systems,'' \emph{{IEEE} Trans. Serv. Comput.}, vol.~8,
  no.~6, pp. 861--873, 2015.

\bibitem{ArtifactsConformance}
D.~Fahland, M.~de~Leoni, B.~F. van Dongen, and W.~M.~P. van~der Aalst,
  ``Behavioral conformance of artifact-centric process models,'' in \emph{{BIS}
  2011}, ser. LNBIP, vol.~87.\hskip 1em plus 0.5em minus 0.4em\relax Springer,
  pp. 37--49.

\bibitem{ArtifactsDecomposedConformance}
------, ``Conformance checking of interacting processes with overlapping
  instances,'' in \emph{{BPM} 2011.}, ser. LNCS, vol. 6896.\hskip 1em plus
  0.5em minus 0.4em\relax Springer, pp. 345--361.

\bibitem{ArtifactsWithRestrictions}
M.~L. van Eck, N.~Sidorova, and W.~M.~P. van~der Aalst, ``Guided interaction
  exploration in artifact-centric process models,'' in \emph{{CBI} 2017}.\hskip
  1em plus 0.5em minus 0.4em\relax {IEEE} Computer Society, pp. 109--118.

\bibitem{DBNetsCPNsAndRelationalDatabase}
M.~Montali and A.~Rivkin, ``Db-nets: On the marriage of colored petri nets and
  relational databases,'' \emph{Trans. Petri Nets Other Model. Concurr.},
  vol.~12, pp. 91--118, 2017.

\bibitem{OCBCFoundation}
W.~M.~P. van~der Aalst, G.~Li, and M.~Montali, ``Object-centric behavioral
  constraints,'' \emph{CoRR}, vol. abs/1703.05740, 2017.

\bibitem{OCBCDiscovery}
G.~Li, R.~M. de~Carvalho, and W.~M.~P. van~der Aalst, ``Automatic discovery of
  object-centric behavioral constraint models,'' in \emph{{BIS} 2017.}, ser.
  LNBIP, vol. 288.\hskip 1em plus 0.5em minus 0.4em\relax Springer, pp. 43--58.

\bibitem{StarStartMVPfromdatabase}
A.~Berti and W.~M.~P. van~der Aalst, ``Extracting multiple viewpoint models
  from relational databases,'' in \emph{{SIMPDA} Revised Selected Papers}, ser.
  LNBIP, vol. 379.\hskip 1em plus 0.5em minus 0.4em\relax Springer, 2019, pp.
  24--51.

\bibitem{FitnessOverview}
A.~F. Syring, N.~Tax, and W.~M.~P. van~der Aalst, ``Evaluating conformance
  measures in process mining using conformance propositions,'' \emph{Trans.
  Petri Nets Other Model. Concurr.}, vol.~14, pp. 192--221, 2019.

\bibitem{PrecisionOverview}
N.~Tax, X.~Lu, N.~Sidorova, D.~Fahland, and W.~M.~P. van~der Aalst, ``The
  imprecisions of precision measures in process mining,'' \emph{Inf. Process.
  Lett.}, vol. 135, pp. 1--8, 2018.

\bibitem{ReplayingHistoryFitnessPrecision}
W.~M.~P. van~der Aalst, A.~Adriansyah, and B.~F. van Dongen, ``Replaying
  history on process models for conformance checking and performance
  analysis,'' \emph{Wiley Interdiscip. Rev. Data Min. Knowl. Discov.}, vol.~2,
  no.~2, pp. 182--192, 2012.

\end{thebibliography}


\begin{thebibliography}{00}
\bibitem{b1} G. Eason, B. Noble, and I. N. Sneddon, ``On certain integrals of Lipschitz-Hankel type involving products of Bessel functions,'' Phil. Trans. Roy. Soc. London, vol. A247, pp. 529--551, April 1955.
\bibitem{b2} J. Clerk Maxwell, A Treatise on Electricity and Magnetism, 3rd ed., vol. 2. Oxford: Clarendon, 1892, pp.68--73.
\bibitem{b3} I. S. Jacobs and C. P. Bean, ``Fine particles, thin films and exchange anisotropy,'' in Magnetism, vol. III, G. T. Rado and H. Suhl, Eds. New York: Academic, 1963, pp. 271--350.
\bibitem{b4} K. Elissa, ``Title of paper if known,'' unpublished.
\bibitem{b5} R. Nicole, ``Title of paper with only first word capitalized,'' J. Name Stand. Abbrev., in press.
\bibitem{b6} Y. Yorozu, M. Hirano, K. Oka, and Y. Tagawa, ``Electron spectroscopy studies on magneto-optical media and plastic substrate interface,'' IEEE Transl. J. Magn. Japan, vol. 2, pp. 740--741, August 1987 [Digests 9th Annual Conf. Magnetics Japan, p. 301, 1982].
\bibitem{b7} M. Young, The Technical Writer's Handbook. Mill Valley, CA: University Science, 1989.
\end{thebibliography}
\iffalse
\fi

\end{document}